\documentclass[graybox]{svmult}

\usepackage{mathptmx}	% selects Times Roman as basic font
\usepackage{helvet}         	% selects Helvetica as sans-serif font
\usepackage{courier}        	% selects Courier as typewriter font
\usepackage{type1cm}	% activate if the above 3 fonts are not available on your system
\usepackage{makeidx} 	% allows index generation
\usepackage{graphicx} 	% standard LaTeX graphics tool when including figure files
\usepackage{multicol}  	% used for the two-column index
\usepackage{multirow}	% used for merging cells
\usepackage{caption}	% used for center caption
\usepackage[bottom]{footmisc} % places footnotes at page bottom
\usepackage{amsmath}

\makeindex     % used for the subject index
\begin{document}

\title*{Designing a belief function-based accessibility indicator to improve web browsing\newline for disabled people}
%\title*{Designing a belief function-based accessibility indicator to browse webpages for disabled people}
\titlerunning{Designing a belief function-based accessibility indicator to browse webpages}
\author{Jean-Christophe Dubois, Yolande Le Gall and Arnaud Martin}
%% Use \authorrunning{Short Title} for an abbreviated version of
%% your contribution title if the original one is too long
\institute{Jean-Christophe Dubois, Yolande Le Gall, Arnaud Martin \at DRUID-IRISA, University of Rennes 1, rue Edouard Branly, 22300 Lannion, FRANCE\\
\email{Jean-Christophe.Dubois@univ-rennes1.fr} \newline 
\email{Yolande.Le-Gall@univ-rennes1.fr}  \newline 
\email{Arnaud.Martin@univ-rennes1.fr}
}


\maketitle

\abstract{The purpose of this study is to provide an accessibility measure of webpages, in order to draw disabled users to the pages that have been designed to be accessible to them. Our approach is based on the theory of belief functions, using data which are supplied by reports produced by automatic web content assessors that test the validity of criteria defined by the WCAG 2.0 guidelines proposed by the World Wide Web Consortium (W3C) organization. These tools detect errors with gradual degrees of certainty and their results do not always converge.  For these reasons, to fuse information coming from the reports, we choose to use an information fusion framework which can take into account the uncertainty and imprecision of information as well as divergences between sources. Our accessibility indicator covers four categories of deficiencies. To validate the theoretical approach in this context, we propose an evaluation completed on a corpus of 100 most visited French news websites, and 2 evaluation tools. The results obtained illustrate the interest of our accessibility indicator.}
%
% Introduction
%
\section{Introduction}
The Web constitutes today an essential source of information and communication. While users have a growing interest in terms of social, cultural and economic value, and in spite of legislations and recommendations of the W3C community for making websites more accessible, its accessibility remains hardly efficient for some disabled or ageing users.  Actually, making websites accessible and usable by disabled people is a challenge \cite{eds10} that society needs to overcome \cite{abascal04}. 

To measure the accessibility of a webpage, several accessibility metrics have been developed \cite{vigo11}. Evaluations are based on the failure to comply with the recommendations of standards, using automatic evaluation tools. They often give a final value, continuous or discrete, to represent content accessibility. However, the fact remains that tests on accessibility criteria are far from being trivial \cite{brajnik04}. Evaluation reports of automatic assessors contain errors considered as certain, but also warnings or potential problems which are uncertain. Moreover there are differences between assessor evaluations, even for errors considered as certain.

This work provides a new measure of accessibility and an information fusion framework to fuse information coming from the reports of automatic assessors allowing search engines to re-rank their results according to an accessibility level, as some users would like \cite{ivory04}. This accessibility indicator considers several categories of deficiencies. Our approach is based on the theory of the belief functions adapted to take into account the defects of accessibility given by several automatic assessors seen as information sources, the uncertainty of their results, as well as the possible conflicts between the sources.

In the sections 2 and 3 we will give a description of accessibility tools based on a recent standard and of data provided in their reports. In the $4^{th}$ section, we will describe the principles of our indicator and develop how we implement the belief functions. In the $5^{th}$ part, we will present an experiment before concluding.
%
% Defect detection of webpage accessibility
%
\section{Defect detection of webpage accessibility}
Various accessibility standards propose recommendations for improving accessibility of webpages. The Web Content Accessibility Guidelines (WCAG 2.0) \cite{brajnik06} proposed by the W3C normalization organization, constitutes an international reference in the field. These guidelines cover a wide range of disabilities (visual, auditory, physical, speech, cognitive, etc.) and several layers of guidance are provided:
\begin{itemize}
\item 4 overall principles: perception, operability, understandability \& robustness; 
\item testable success criteria: for each guideline, testable success criteria are provided. Every criterion is associated to one of the 3 defined conformance levels (A, AA and AAA), each representing a requirement of accessibility for users. 
\end{itemize}
Several automatic accessibility assessors, based on various accessibility standards, have been developed \cite{caldwell08} for IT professionals. Their limits depend on the automatic tests. Because it is at present not possible to test some criteria about the quality of some pages, some assessor results are given with ambiguity. Consequently, the existing automatic assessors look for the criteria which are not met and give the defects according to 3 levels of validity: the number of errors, which are estimated certain, the number of likely problems (warnings) whose reality is not guaranteed and the number of potential problems (also called generic or non testable) which leads to a complete uncertainty on the tested criterion accessibility.

Finally, even though the results obtained by different assessors match for some tested common criteria, results can differ, even for errors considered as certain.
%
% Proposed accessibility indicator
%
\section{Proposed accessibility indicator}
After a request, the indicator has to supply information describing to users the accessibility level of each webpage proposed by a search engine. Presented simultaneously with these pages, the indicators' information cover two aspects:
\begin{itemize}
\item the accessibility for categories of deficiencies: as previously proposed for accessibility estimation \cite{dempster67} we use 4 major categories: visual, hearing, motor and cognitive deficiencies, as defined by \cite{vanderheiden91}. They are called  ``deficiency frames''; 
\item the level of accessibility for each deficiency frame.
\end{itemize}
Collecting results from several assessors has allowed us to benefit from each of their performance. In addition, it strengthens accessibility evaluation for similar results and manages conflicts in case of disagreements. Automatic assessors check a set of criteria which correspond to many deficiencies. As our accessibility evaluation varies for every deficiency frame, our method consists in selecting the relevant criteria for each deficiency frame and then balancing each criterion to consider the difficulties met by users in case of failure. This weighting is based on the criterion conformance level (A, AA, AAA), which corresponds to decreasing priorities (A: most important, etc.). The errors and problems detected for every criterion of the accessibility standard affect the accessibility indicator of the Web content tested according to the deficiency frame the criterion belongs to, its weighting within the frame, the number of occurrences when it is analyzed as a defect in the webpage and the defect's degree of certainty (error, likely or potential problem).
%
% Defect detection and accessibility evaluation
%
\section{Defect detection and accessibility evaluation}
After collecting webpage Uniform Resource Locators $(URL_p)$ selected by a search engine from a request, these addresses are supplied to the accessibility assessors and successively for each page, we detect accessibility defects, then estimate accessibility level by deficiency frame for each assessor, before fusing the data by deficiency frame and taking the decision for every deficiency frame \cite{dubois14}.
\subsection{Assessor evaluations of selected pages}
Each $URL_p$ is submitted to the accessibility evaluation tests by each assessor \textit{i} that tests all the criteria \textit{k} of the WCAG 2.0 standard, and the following data are collected by a filter that extracts the required data for each deficiency frame:
\begin{itemize}
\item $N_{k,i}^e$ : errors observed for a criterion \textit{k} by an assessor \textit{i};
\item $N_{k,i}^c$  : correct checkpoints for a criterion \textit{k}  by an assessor \textit{i};
\item $T_{k,i}^e$  : tests that can induce errors for a criterion \textit{k}  by an assessor \textit{i};
\item $N_{k,i}^l$  : likely problems detected for a criterion \textit{k}  by an assessor \textit{i};
\item $T_{k,i}^l$  : tests that can induce likely problems for a criterion \textit{k}  by an assessor \textit{i};
\item $N_{k,i}^p$  : potential problems suspected for a criterion \textit{k}  by an assessor \textit{i};
\item $T_{k,i}^p$  : tests that can induce potential problems for a criterion \textit{k}  by an assessor \textit{i};
\item $T_i$   : total tests by an assessor \textit{i}, with:
\begin{equation}
T_i=\sum_k (N_{k,i}^e+N_{k,i}^l+N_{k,i}^p+N_{k,i}^c)
\end{equation}
\end{itemize}
\subsection{Accessibility indicator level of the pages}
To model initial information including uncertainties, the reliability of the assessors seen as information sources and their possible conflicts, we use the theory of belief functions \cite{brajnik09} \cite{shafer76}. Our objective is to define if a webpage is accessible \textit{(Ac)} or not accessible $(\overline{Ac})$ and to supply an indication by deficiency frame. Consequently, these questions can be handled independently for every deficiency frame $\Omega_h= {\{Ac,\overline{Ac}\}}$. We can consider every power set $2^{\Omega_h}= \{\emptyset,Ac,\overline{Ac},\Omega\}$. 

The estimation of the accessibility \textit{Ac}  for a deficiency frame \textit{h} and a source \textit{i} (assessor) is estimated from the number of correct tests for each of the criteria \textit{k} occurring in this frame, and from their conformance level represented by $\alpha_k$: 
\begin{equation}
E(Ac)_{h,i} = \frac{\sum_k (N_{k,i}^c * \alpha_k)}{T_i}
\end{equation}

The estimation of the non accessibility $\overline{Ac}$ for a deficiency frame \textit{h} and a source \textit{i} is estimated from the number of errors for each of the criteria \textit{k} occurring in this frame, and from the $\alpha_k$  coefficient. A weakening $\beta_i^e$ coefficient is also introduced to model the degree of certainty of the error:
\begin{equation}
E(\overline{Ac})_{h,i} = \frac{\sum_k (N_{k,i}^e * \alpha_k * \beta_i^e)}{T_{k,i}^e}
\end{equation}
The estimation of the ignorance $\Omega_h$  for a deficiency frame \textit{h} and a source \textit{i} is estimated from the number of likely and potential problem for each of the criteria k occurring in this frame, and from the $\alpha_k$ coefficient. The weakening coefficients $\beta_i^l$  or $\beta_i^p$  are also used to model the degree of certainty of the problem:                                                                       
\begin{equation}
E(\Omega_{h,i} = \frac{\sum_k (N_{k,i}^l * \alpha_k * \beta_i^l +N_{k,i}^p * \alpha_k * \beta_i^p)} {\sum_k (T_{k,i}^l + T_{k,i}^p)}
\end{equation}
The mass functions of the subsets of $2^{\Omega_h}$ are computed from the estimations:
\begin{equation}
m(Ac)_{h,i} = \frac{E(Ac)_{h,i}} {E(Ac)_{h,i} + E(\overline{Ac})_{h,i} + E(\Omega)_{h,i}}
\end{equation}
\begin{equation}
m(\overline{Ac})_{h,i} = \frac{E(\overline{Ac})_{h,i}} {E(Ac)_{h,i} + E(\overline{Ac})_{h,i} + E(\Omega)_{h,i}}
\end{equation}
\begin{equation}
m(\Omega)_{h,i} = \frac{E(\Omega)_{h,i}} {E(Ac)_{h,i} + E(\overline{Ac})_{h,i} + E(\Omega)_{h,i}}
\end{equation}
In addition, the source reliability can be modeled \cite{martin08} with a $\delta_i$ coefficient, which constitutes a benefit when some assessors are more efficient than others:
%\begin{equation}
%m^{\delta_i}(Ac)_{h,i} = \delta_i * m(Ac)_{h,i}
%\hspace{2cm}
%m^{\delta_i}(\overline{Ac})_{h,i} = \delta_i * m(\overline{Ac})_{h,i}
%\end{equation}
%$$m^{\delta_i}(\Omega)_{h,i} = 1 - \delta_i * (1 - m(\Omega)_{h,i}) $$
\begin{eqnarray}
\label{Trio}
\left\{
\begin{array} {rcl}
\vspace{0.2cm}
m^{\delta_i}(Ac)_{h,i} = \delta_i * m(Ac)_{h,i} \\
\vspace{0.2cm}
m^{\delta_i}(\overline{Ac})_{h,i} = \delta_i * m(\overline{Ac})_{h,i} \\
\vspace{0.2cm}
m^{\delta_i}(\Omega)_{h,i} = 1 - \delta_i * (1 - m(\Omega)_{h,i})
\end{array}
\right.
\end{eqnarray}
                                                
\subsection{Merging assessor results and decision-making}
Once the masses for each assessor have been obtained, a fusion of the results is conducted by deficiency frame, using the conjunctive rule \cite{smets93}, to combine them and give information in the form of a mass function. These rule properties, which strengthen common results and manage conflicts between sources, are particularly relevant in this context, to deal with divergences between assessor evaluations. To calculate the final decision $D_h (URL_p)$ for a page by deficiency frame, we use the pignistic probability \cite{smets93}. 

There are several ways of presenting the accessibility indicator to users. To visualize the deficiency frames, existing specific pictograms are effective. To present the accessibility level we discretize the decision into 5 levels (very good, good, moderate, bad or very bad accessibility) using thresholds and visualized it by an "arrow":
\begin{itemize}
\item if $D_h<S_1$, the Web content accessibility is very bad ($\downarrow$),
\item if $S_1<D_h<S_2$, the Web content accessibility is bad ($\searrow$),
\item if $S_2<D_h<S_3$, the Web content accessibility is moderate ($\rightarrow$),
\item if $S_3<D_h<S_4$, the Web content accessibility is good ($\nearrow$),
\item if $S_4<D_h$, the Web content accessibility is very good ($\uparrow$). 
\end{itemize}
%
% Experiments
%
\section{Experiments}
To validate our approach, we present here the results obtained on a set of 100 news Websites, among the most visited ones, all referenced by the OJD \footnote{OJD: http://www.ojd.com/Chiffres/Le-Numerique/Sites-Web/Sites-Web-GP} organization which provides certification and publication of attendance figures for websites. We test their homepages, following a study \cite{nielsen01} concluding that their usability is predictive of the whole site. We chose two open source assessors AChecker, (source 1) \cite{gay10}, and TAW (source 2) from which we extract automatically the accessibility test results. Weight and threshold values given in Table \ref{tabconstantes} were previously empirically defined from Webpages \footnote{Sites labeled by Accessiweb: http://www.accessiweb.org/index.php/galerie.html} assumed to be accessible.

\vspace{-0.3cm}
\begin{table}
\centering
\begin{tabular}{|c|c|c|c|}
\hline 
 \multirow{3}{*}{Weightings} & $\alpha_1$ ; $\alpha_2$ ;$\alpha_3$ & A, AA, AAA conformance levels & 1 ; 0.8 ; 0.6 \\
& $\beta_i^e$ ; $\beta_i^l$ ;$\beta_i^p$ & Certainty levels of errors or problems & 1 ; 0.5 ; 1 \\
& $\delta_1$ ; $\delta_2$ ;$\alpha_3$ & AChecker and TAW reliabilities (sources) & 1 ; 1\\
\hline
Thresholds & S1 ; S2 ; S3 ; S4 & Accessibility indicator levels & 0.6 ; 0.7 ; 0.8 ; 0.9 \\
\hline 
\end{tabular}
\caption{\label{tabconstantes}Constant values for our accessibility metric.}
\end{table}

\begin{figure}
\centering
\includegraphics[height=4.8cm]{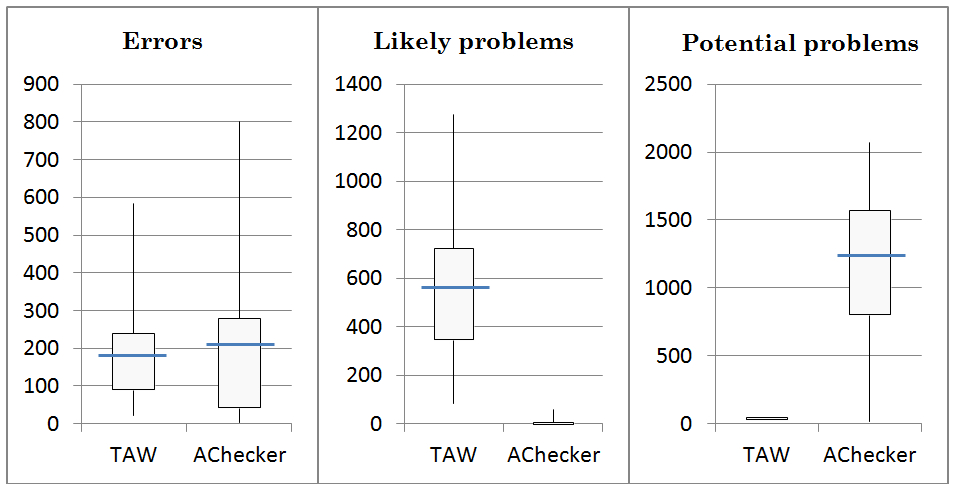}
\caption{\label{errors}Results of automatic assessors.}
\end{figure}

\vspace{-0.4cm}
The results of these sources are summarized in Figure~\ref{errors} for the 3 levels of certainty defects. The box plots present how  their  defects  are distributed: minimum and maximum (whiskers), $1^{st}$ (bottom box plot) and $3^{rd}$ quartiles (top box plot) and average (horizontal line). We observe similarities between the assessors' results for the errors detected as certain, but also huge differences for the likely (warnings) and potential (non testable) problems. The number of likely problems is almost null for AChecker and the potential one remains always the same for TAW.

\vspace{-0.3cm}

\begin{figure}
\centering
\includegraphics[height=5.4cm]{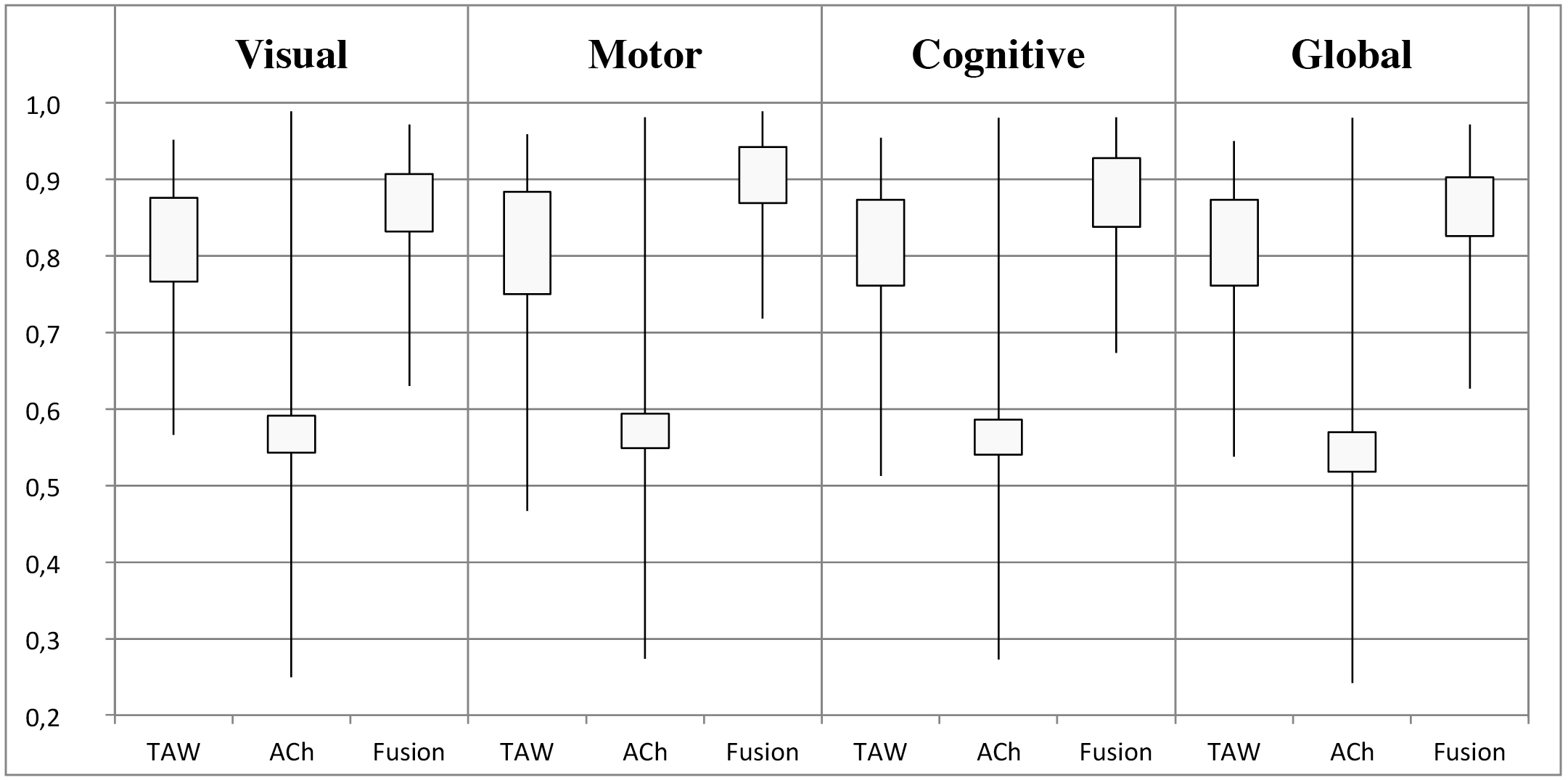}
\caption{\label{handicaps}Accessibility indicator results.}
\end{figure}

\vspace{-0.4cm}
The detected defects are taken into account in our accessibility indicator results presented in Figure~\ref{handicaps}. The mass function values of accessibility \textit{m(Ac)} for the 2 sources, TAW and AChecker, and the fusion result are visualized for 3 deficiency frames among the 4, and globally for all deficiencies.  Firstly, we can see that \textit{m(Ac)} is not evenly distributed between the 2 sources: their distributions of errors (Figure~\ref{handicaps}) are comparable even if there is a larger range for AChecker; however the mass function of accessibility is smaller for AChecker compared to TAW. This is due to the more numerous potential problems (non testable criteria) detected by the AChecker assessor, increasing substantially the denominator in the computation of \textit{m(Ac)} (Eq.~5). By the way, the values of $E(\Omega)$ and consequently of $m(\Omega)$, are more important, as the $\beta_i^p$ weight for potential problems is 2 times higher than $\beta_i^l$ for the likely problems (warnings). We can also notice that the fusion result obtained by the conjunctive rule strengthens the mass functions of the 2 assessors.

\vspace{-0.4cm}

\vspace{-0.5cm}
In this corpus, visual and cognitive deficiencies have a higher impact on content accessibility than the motor ones. This is logical for news websites, as their homepages include a large number of images. By the way, the motor indicator is less impacted, in particular by the lack of alternatives for images, useful for visual and cognitive deficiencies. Finally, we observe a similarity between the visual and global indicators, as around 80\% of all the checkpoints concern visual deficiencies and also because these controls are properly taken into account by assessors.

\vspace{-0.4cm}
\begin{table}
\begin{center}
\begin{tabular}{|p{3cm}|p{1.5cm}|p{1.5cm}|p{1.5cm}|p{1.5cm}|}
\hline 
 \multirow{2}{*}{Web content ($URL_p$)} & \multicolumn{4}{c|}{Decision} \\
& Visual & Motor & Cognitive & Global \\
\hline
LeParisien.fr 	& 0.972  $\uparrow$		& 0.989  $\uparrow$		& 0.974  $\uparrow$		& 0.971 $\uparrow$ \\
Famili.fr		& 0.769  $\rightarrow$	& 0.924  $\uparrow$		& 0.838  $\nearrow$		& 0.766 $\nearrow$\\
Arte.tv		& 0.701  $\rightarrow$	& 0.718  $\rightarrow$	& 0.717  $\rightarrow$	& 0.686 $\searrow$\\
 LePoint.fr		& 0.630  $\searrow$		& 0.725  $\rightarrow$	& 0.673  $\searrow$		& 0.627 $\searrow$\\
\hline 
\end{tabular}
\caption{\label{tabresults}Examples of detailed accessibility results by deficiency frame.}
\end{center}
\end{table}

\vspace{-0.7cm}
In Table~\ref{tabresults} are presented detailed results for several sites with significant indicator result differences. For examples, \textit{LePoint.fr} and \textit{Arte.tv}, respectively $19^{th}$ and $33^{th}$ most consulted websites in France, obtain only 0.627 and 0.686 for the global result, whereas \textit{LeParisien.fr}, ranked $12^{th}$, reaches 0.971. For \textit{Family.fr} we observe differences between the deficiencies, nevertheless focus on accessibility generally benefits all deficiencies on the whole corpus.
%
% Conclusion
%
\section{Conclusion}
We present an indicator estimating webpage accessibility levels for distinct categories of deficiencies, in order to supply easily understandable accessibility information to users on pages proposed by a search engine. Our method based on belief function theory fuses results from several automatic assessors and considers their uncertainties. An accurate modelization of the assessor characteristics and of the impact of defect guideline criteria on accessibility is proposed. An experiment performed on a set of 100 news websites validates the method, which benefits from each of the assessor performances on specific criterion tests. Our future research will focus on the implementation of a user's personal weighting to balance the importance of criteria.
%
% Références
%
\vspace{-0.3cm}
%%%%%%%%%%%%%%%%%%%%%%%% referenc.tex %%%%%%%%%%%%%%%%%%%%%%%%%%%%
% 												
%											References
% 
%%%%%%%%%%%%%%%%%%%%%%%% Springer-Verlag %%%%%%%%%%%%%%%%%%%%%%%%%%

\end{document}